\documentclass[pss,fleqn]{w-art}
\usepackage{times}
\usepackage{w-thm}
\usepackage{graphicx}
\begin{document}
\newcommand {\bea}{\begin{eqnarray}}
\newcommand {\eea}{\end{eqnarray}}
\newcommand {\bb}{\bibitem}
\DOIsuffix{theDOIsuffix}
\Volume{XX}
\Issue{1}
\Month{01}
\Year{2003}
\pagespan{1}{}
\Receiveddate{}
\Reviseddate{}
\Accepteddate{}
\Dateposted{}
\title{New World of Gossamer Superconductivity}

\author[K. Maki]{Kazumi Maki\inst{1,2}}
\address[\inst{1}]{Max-Planck-Institut f\"{u}r Physik komplexer Systeme,
N\"{o}thnitzer Str. 38, D-01187, Dresden, Germany}
\author[S. Haas]{Stephan Haas\inst{2}}
\address[\inst{2}]{Department of Physics and Astronomy, University of Southern
California, Los Angeles, CA 90089-0484 USA}
\author[D. Parker]{David Parker\inst{2}}

\author[H. Won]{Hyekyung Won\inst{1,3}} 
\address[\inst{3}] {Department of Physics, Hallym 
University, Chuncheon 200-702, South Korea}
\author[B.Dora]{Balazs Dora\inst{4}}
\address[\inst{4}]{Department of Physics, Budapest University of
Technology and Economics H-1521,Budapest, Hungary}
\author[A. Virosztek]{Attila Virosztek\inst{4}}


\begin{abstract}
Since the discovery of the high-T$_{c}$ cuprate superconductor 
La$_{2-x}$BaCuO$_{4}$ in 1986 by Bednorz and M\"{u}ller, controversy regarding
the nature or origin of this remarkable superconductivity has continued.
However, d-wave superconductivity in the hole-doped cuprates,
arising due to the anti-paramagnon exchange, was established around 1994.
More recently we have shown that the mean field theory, like the BCS theory
of superconductivity and Landau's Fermi liquid theory are adequate to 
describe the cuprates.  The keys for this development are the facts that a)the 
pseudogap phase is d-wave density wave (dDW) and that the high-T$_{c}$ cuprate
superconductivity is gossamer (i.e. it exists in the presence of dDW).
\end{abstract}
\maketitle                   





\section{Introduction}

The epoch-making discovery of high-T$_{c}$ cuprate superconductivity
by Bednorz and M\"uller \cite{1} put the entire superconductivity community in exaltation and confusion.  This situation is nicely described by Enz \cite{2}.
As to the theoretical modeling of high-T$_{c}$ cuprates
the most influential proposal was the two-dimensional one band
Hubbard model and related resonance valence band state proposed by P.W.
Anderson \cite{3}.  In particular, Anderson gave the ground state wave function
as 
\bea
\Phi &=& \prod_{i}(1-d_{i})|BCS>
\eea
where $|BCS>$ is the BCS wave function for s-wave superconductors \cite{4} and 
$\prod(1-d_{i})$ is the Gutzwiller projector where $d_{i}= n_{i\uparrow}
n_{i\downarrow}$.  This Gutzwiller operator annihilates all the doubly
occupied states.  Then in 1994 the d-wave symmetry of high-T$_{c}$ cuprate
superconductivity was established through Josephson interferometry \cite{5,6}
and powerful angle-resolved photoemission spectroscopy (ARPES) \cite{7}.
Therefore, at a minimum $|BCS>$ in Eq.(1) has to be replaced by the corresponding one for 
d-wave superconductors \cite{8,9}.

In 2002 R.B. Laughlin proposed that the 
Gutzwiller operator used by Anderson should be replaced by the more
general Jastrow function, since the Gutzwiller projector
is intractable \cite{10}.  But it is well known that both the d-wave
superconducting wave function and the d-wave density wave (dDW) wave functions 
can be recast in the Jastrow form \cite{11}.  

We show in Fig. 1 the schematic phase diagram of the hole-doped high-T$_{c}$
cuprates.  The antiferromagnetic state at zero doping (x=0) vanishes
around x=3\%.  Also the d-wave superconducting dome appears for $5 \leq x
\leq 25$.  Under T* there is the pseudogap phase.  
\begin{figure}[h]
\includegraphics[width=8cm]{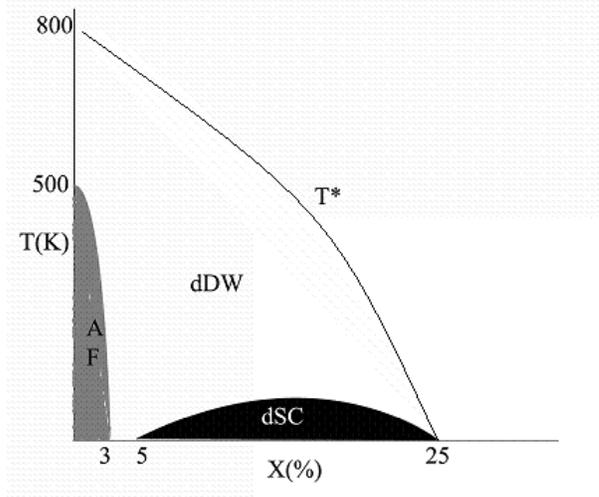}
\caption{The phase diagram for the high-T$_{c}$ cuprates}
\end{figure}
Recently a few people proposed that the pseudogap phase is d-wave density
wave (dDW) \cite{12,13,14,15}.  Indeed the giant Nernst effect observed
in the underdoped region in LSCO, YBCO and Bi-2212 \cite{16,17,18,19,20}
and the angle-dependent magnetoresistance in 
Y$_{0.68}$Pr$_{0.32}$Ba$_{2}$Cu$_{3}$O$_{7}$ can be described consistently
in terms of dDW \cite{21,22}.  If we accept that the pseudogap phase is dDW,
we expect that d-wave superconductivity coexists with dDW.  In other words
the superconductivity in high-T$_c$ cuprates is gossamer superconductivity
\cite{23,24}.

In 1993 Volovik \cite{25} showed how to calculate 
the quasiparticle density of states
in the vortex state in d-wave superconductors.  The striking $\sqrt{H}$
dependence of the specific heat has been observed in single crystals
of YBCO \cite{26,27}, LSCO \cite{28}, and Sr$_{2}$RuO$_{4}$ \cite{29,30},
where H is the magnetic field strength.  This
analysis has been extended into several
directions: a) thermodynamic functions; b) thermal conductivity
c) scaling relations; and d) for a variety of gap
functions $\Delta({\bf k})'s$ \cite{31,32,33,35,36,sal}.  Unfortunately 
the literature on Volovik's effect is rather confused.  We recommend that
readers study Ref. \cite{sal} for a brief summary.  Since 2001
Izawa et al have succeeded in determining the superconducting gap functions
$\Delta({\bf k})'s$ in Sr$_{2}$RuO$_{4}$ \cite{37}, CeCoIn$_{5}$ \cite{38},
$\kappa$-(ET)$_{2}$Cu(NCS)$_{2}$ \cite{39}, YNi$_{2}$B$_{2}$C \cite{40},
PrOs$_{4}$Sb$_{12}$ \cite{41,42} and UPd$_{2}$Al$_{3}$ \cite{43,44}
through the angle-dependent magnetothermal conductivity (ADMTC).  
These are shown in Fig. 2.
\begin{figure}[h]
\includegraphics[width=4cm]{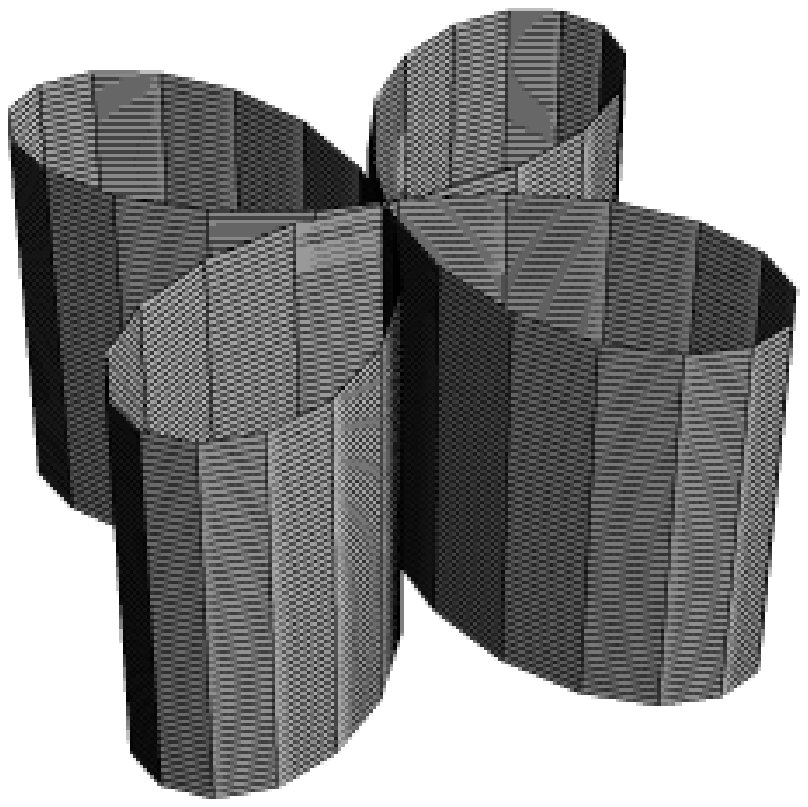}
\includegraphics[width=2cm]{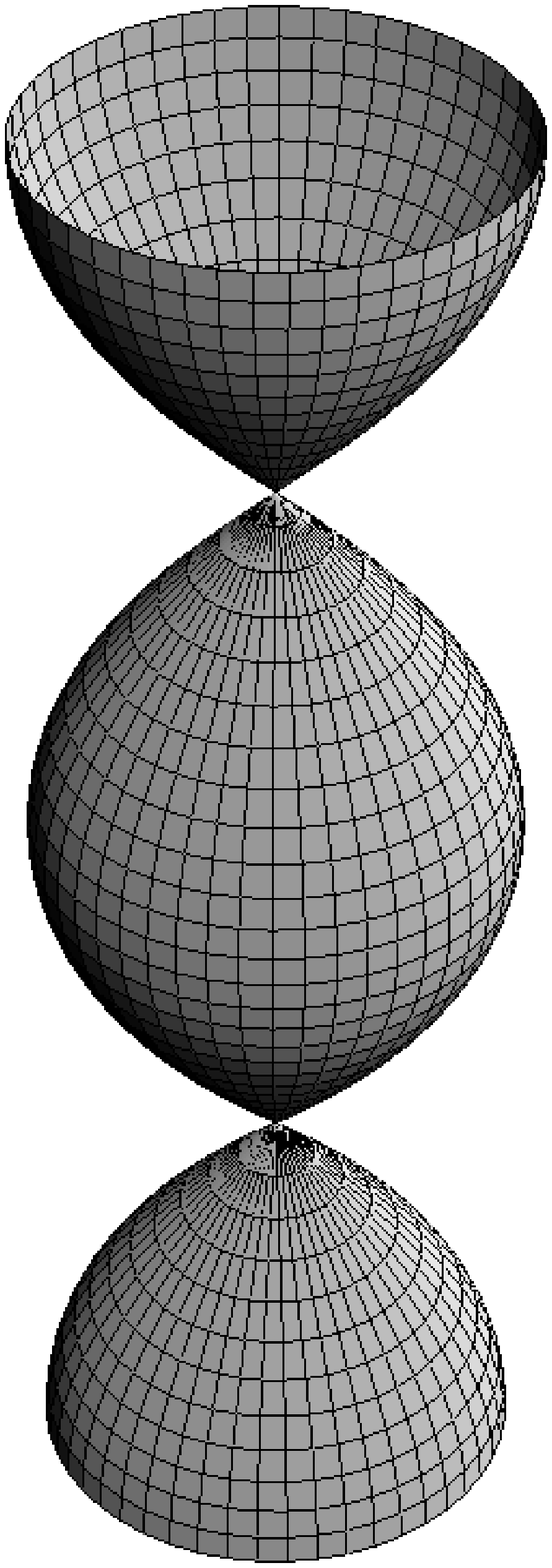}
\includegraphics[width=2.3cm]{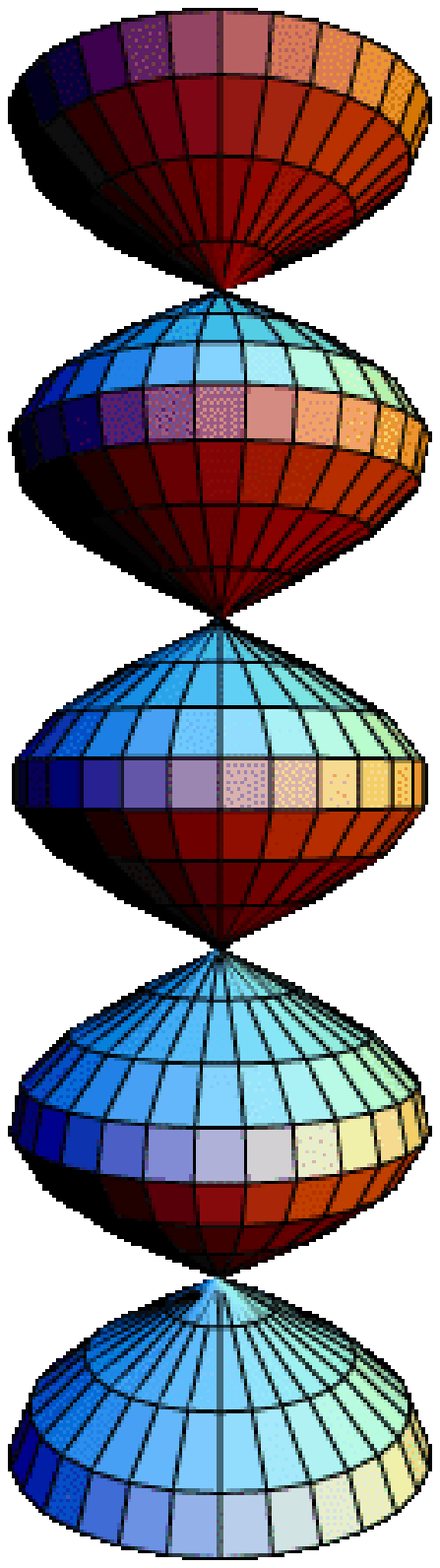}
\includegraphics[width=4cm]{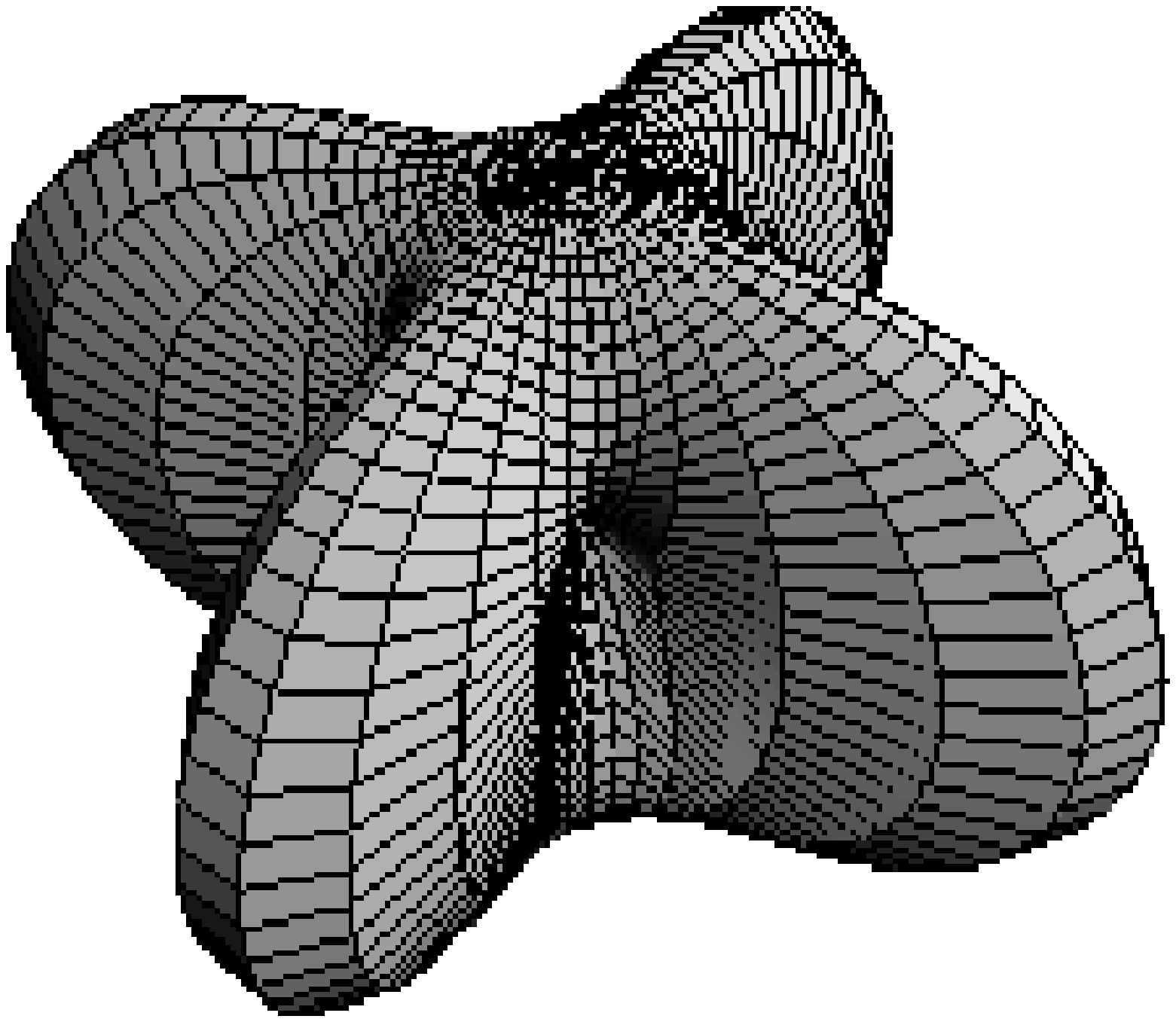}
\includegraphics[width=4cm]{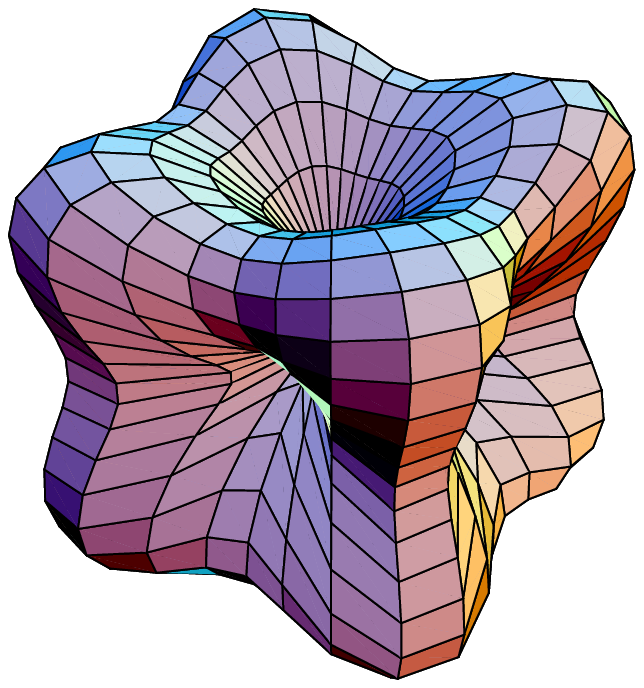}
\includegraphics[width=4cm]{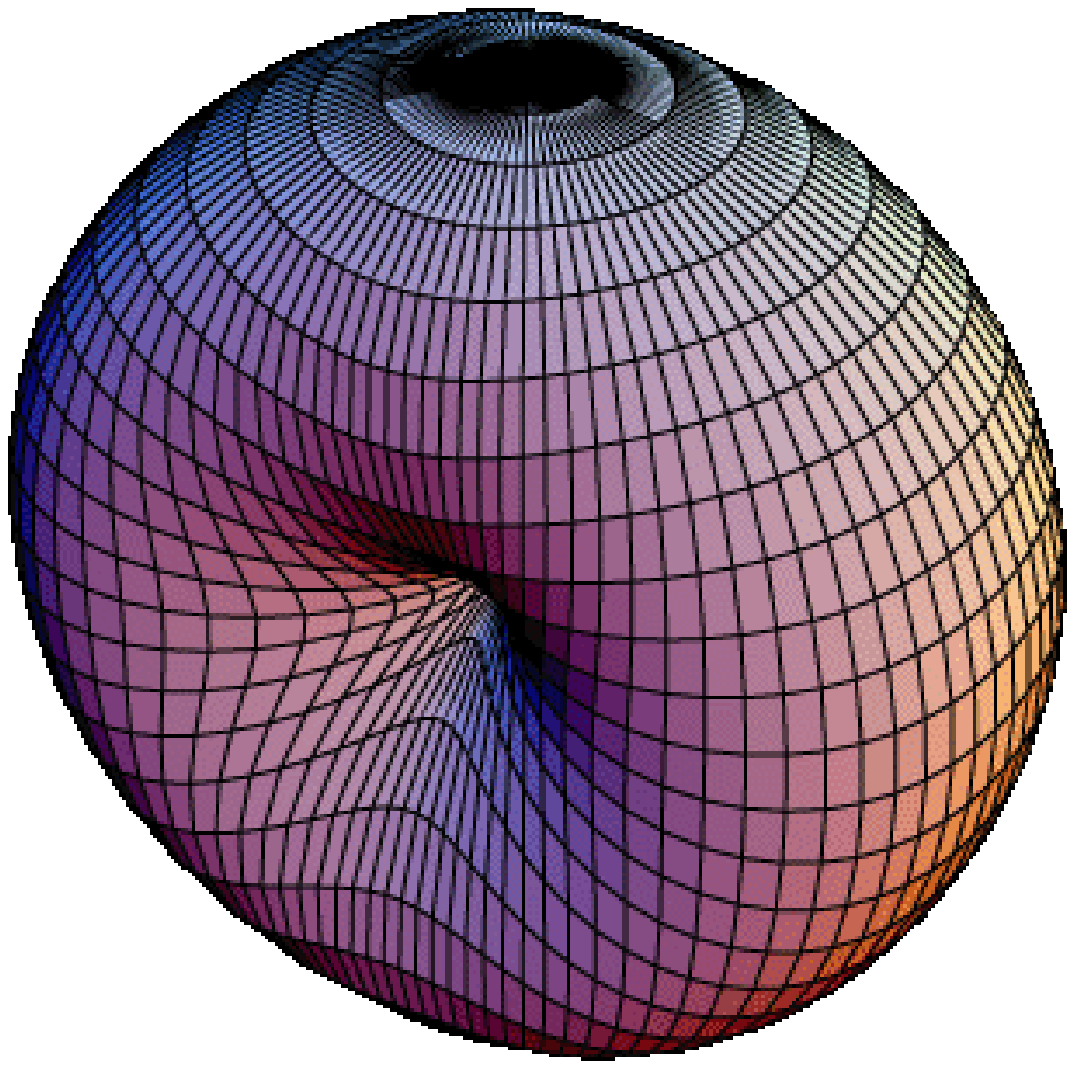}
\caption{Order parameters from top left: d-wave - high-T$_{c}$ cuprates,
CeCoIn$_{5}$, $\kappa$-(ET)$_{2}$Cu(NCS)$_{2}$; chiral 
f-wave - Sr$_{2}$RuO$_{4}$;
g-wave - UPd$_{2}$Al$_{3}$; s+g-wave - YNi$_{2}$B$_{2}$C; p+h-wave, 
PrOs$_{4}$Sb$_{12}$ A-phase;  p+h-wave, PrOs$_{4}$Sb$_{12}$ B-phase.}
\end{figure}
These experiments are only possible due to a) availability of high-quality
single crystals with $ RRR > 100$ b) low-temperature facility operating
at 1000 - 10 mK, and c) the recent theoretical development \cite{sal}. 

In Fig. 3 we show the phase diagram of the hole-doped high-T$_{c}$ cuprates
\cite{24}.  As you may recognize, we have replaced
\begin{figure}[h]
\includegraphics[width=8cm,angle=270]{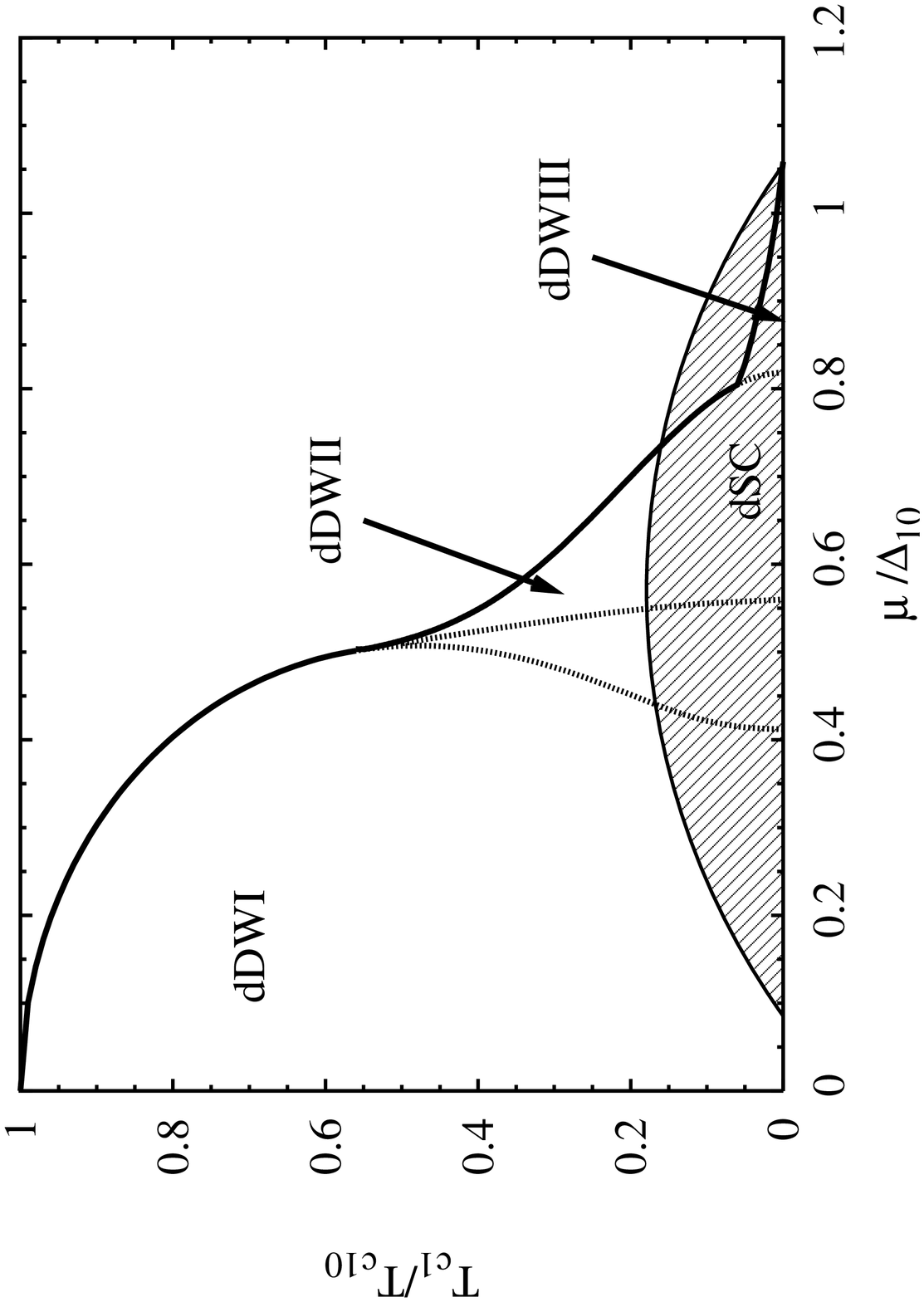}
\caption{The phase diagram for the high-T$_{c}$ cuprates}
\end{figure}
the pseudogap phase with d-wave density wave (dDW).  However, instead of our
usual phase diagram, we take the chemical potential $\mu$
as a control parameter and show that dDW exists in three varieties.

Also the
T$_{c}$ of dDW T$_{c1}$ (= T*) is determined by
\bea
-\ln(\frac{T_{c1}}{T_{c10}}) &=& Re \Psi(\frac{1}{2}-\frac{i\mu}{2\pi 
T_{c1}})-\Psi(\frac{1}{2})
\eea
where T$_{c10} \simeq 800 K$ is the T$_{c1}$ in the limit $\mu=0$.
Here $\mu$ is the chemical
potential and $\Psi(z)$ is the di-gamma function.  Eq (1) is the same as for
the s-wave or d-wave superconductors in the presence of the Pauli
paramagnetic term \cite{60,61}.  Also as shown by the broken curve T$_{c1}$
in Eq. (1) bends back for $T_{c1}/T_{c10} \leq 0.55 $.  However, if you
allow a spatial variation for dDW like $\Delta({\bf r}) \sim \cos({\bf q}
\cdot {\bf r})$ we have to solve a new equation 
\bea
-\ln(\frac{T_{c1}}{T_{c10}}) &=& Re \langle ( 1 \pm \cos(2\phi))
\Psi(\frac{1}{2} - \frac{i\mu(1-p\cos\phi)}{2\pi T}) \rangle
- \Psi(\frac{1}{2})
\eea
where $p= v|q|/2\mu$ and is determined to optimize T$_{c1}$.
Actually Eq. (2) is the same as for the Fulde-Ferrell-Larkin-Ovchinnikov
state in d-wave superconductors, as discussed in \cite{62}, and its solution
is known.  According to \cite{62} dDW splits further into dDW II and dDW III.
In the region dDW II and dDW III
we find ${\bf q} \parallel$ [110] for the - sign in Eq. (3)
and $\parallel$ [100] for the + sign, respectively.
Here $<\ldots>$ means the average
over $\phi$.  Therefore it appears that together with periodic dDW's
we can reproduce the observed T$_{c1}$ for dDW.  Also the theoretical study
of dDW II and dDW III will be of great interest.

\section{D-wave density waves}

As already mentioned many people have proposed d-wave density wave (dDW)
\cite{12,13,14,15} 
for the pseudogap phase of high-T$_{c}$ cuprate superconductors.  But until
recently no quantitative test of these  proposals was available.  Recently we
have shown that the giant Nernst effect observed in the pseudogap phase in
YBCO, LSCO and Bi-2212 \cite{16,17,18,19} can be described in terms
of dDW \cite{20}.  First of all, we stress that the dDW in the pseudogap
phase is very different from that proposed in \cite{12,13,14,15}.  The
present dDW is incommensurate and possesses the U(1) gauge symmetry while
the earlier proposal is the descendant of the flux phase or the staggered
phase \cite{45,46} and carries miniscule loop currents which have not been
observed.  Furthermore, it is clear that such commensurate dDW with Z$_{2}$
symmetry are unstable
in the 3D environment, and cannot have the chemical potential as a control
parameter.

The quasiparticle energy of dDW is given by  \cite{22}
\bea
E({\bf k}) &=& \pm \sqrt{\xi^{2}(k)+\Delta^{2}\sin^{2}(2\phi)} - \mu
\eea
with 
\bea
\xi({\bf k}) = v(k_{\parallel}-k_{F})+\frac{v^{'}}{c}\cos(ck_{z})
\eea
where $k_{\parallel}$ is the radial component in the x-y plane and
v and v$^{'}$ are the Fermi velocities and $\tan(\phi) = k_{y}/k_{x}$.
As we shall see the chemical potential plays the crucial role in the
construction of the phase diagram of high-T$_{c}$ cuprate superconductors.
But the chemical potential is absent in the descendant of the staggered 
phase as in \cite{47} for example.  Indeed as already discussed in
\cite{24} such a model lacks physical relevance.  Then the quasiparticle
density of states is given by
\bea
N(E)/N_{0} &=& G(x-y)
\eea
where 
\bea
G(x)&=& \frac{2x}{\pi}K(x)\,\,\,\mathrm{for\,\, x \leq 1} \\
    &=& \frac{2}{\pi}K(x^{-1})\,\,\,\mathrm{for\,\, x > 1}.
\eea
\begin{figure}[h!]
\includegraphics[width=7cm,angle=270]{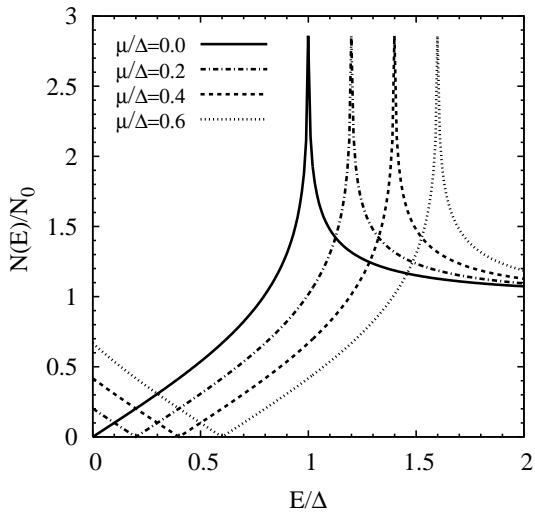}
\caption{The quasiparticle density of states for a dDW superconductor}
\end{figure}
and $x=E/\Delta$, $y= \mu/\Delta$ and K(x) is the complete elliptic
integral.  $N(E)/N_{0}$ for a few$\mu$'s  is shown in Fig. 3.  
As we shall see later $\mu \sim \Delta_{2}$ is essential for the 
presence of gossamer superconductivity.  In addition, the presence of
nonzero $\mu$ is required to account for the Fermi arcs seen in ARPES 
\cite{48}.

Therefore how to identify d-DW or more generally unconventional density
wave (UDW) is the central issue \cite{53}.  As is well known \cite{49,50},
in a magnetic field the quasiparticle energy in UDW is quantized \`{a} la 
Landau.  In other words, except for the n=0 Landau level, the quasiparticle
energy gap $ \sim \sqrt{|B|}$ opens up, 
where B is the strength of the magnetic
field.  This leads to the giant Nernst effect \cite{20}, the angle-dependent
magnetoresistance (ADMR) \cite{51,52} and the nonlinear Hall effect \cite{22}.

From the experimental data of ADMR we have identified UDW in the low
temperature phase (LTP) in $\alpha$-(ET)$_{2}$KHg(SCN)$_{4}$ \cite{51}, in the
metallic phase of (TMTSF)$_{2}$X with X= PF$_{6}$ and ReO$_{4}$ \cite{52,57}
both under pressure and magnetic fields.  More recently we have identified
dDW in the pseudogap phases in the underdoped high-T$_{c}$ cuprates
Y$_{0.86}$Pr$_{0.32}$CuO$_{4}$ with T$_{c}$ = 55 K \cite{21}, and the
heavy-fermion system CeCoIn$_{5}$ \cite{55}.  Also it will be of great
help to explore both the giant Nernst effect and the nonlinear Hall effect
in these systems.

\section{Concluding Remarks}

Earlier we have seen that most of the metallic ground states in
high-T$_{c}$ cuprates, heavy-fermion conductors and organic conductors
belong to one of the mean field ground states: a) unconventional 
superconductivity, b) unconventional density wave; or c) the coexistence
of both unconventional superconductivity and UDW.  
The present analysis suggests
that a) most of the the so-called ``non-Fermi liquid'' state is in fact the
Fermi liquid \`{a} la Landau and UDW, b) the superconductivity 
in both high-T$_{c}$ cuprates
and CeCoIn$_{5}$ are gossamer; and c) the superconductivity in 
$\kappa$-(ET)$_{2}$ salts and in Bechgaard salts (TMTSF)$_{2}$PF$_{6}$
and URu$_{2}$Si$_{2}$ are also gossamer \cite{66,67}.  This suggests 
a vast forest of gossamer superconductors are awaiting our exploration.
                                                                                \begin{acknowledgement}
We have benefitted from collaborations and discussions
with Carmen Almasan, Christian Bernhard,
Tao Hu, Bernhard Keimer, Viorel Sandu, Manfred Sigrist, and Peter
Thalmeier.
\end{acknowledgement}


\begin{thebibliography}{99}
\bb{1} J.G. Bednorz and K.A. M\"uller, Z. Phys. B {\bf 64}, 180 (1986).
\bb{2} C.P. Enz, A Course in Many-Body Theory Applied to Solid State Physics
(World Scientific, Singapore, 1992).
\bb{3} P.W. Anderson, Science {\bf 235}, 1196 (1987); The Theory of 
High-T$_{c}$ Superconductivity (Princeton, 1998).
\bb{4} J. Bardeen, L.N. Cooper and J.R. Schrieffer, Phys. Rev. {\bf 108},
1185 (1957).
\bb{5} D.J. Van Harlingen, Rev. Mod. Phys. {\bf 67}, 515 (1995).
\bb{6} C.C. Tsuei and J.R. Kirtley, Rev. Mod. Phys. {\bf 72}, 969 (2000).
\bb{7} A. Damascelli, Z. Houssain and Z.X. Shen, Rev. Mod. Phys. {\bf 72},
969 (2000).
\bb{8} H. Won and K. Maki, Phys. Rev. B {\bf 49}, 1397 (1994).
\bb{9} K. Maki and H. Won, Phys. Rev. Lett. {\bf 72}, 1758 (1994).
\bb{10} R.B. Laughlin, cond-mat/0209269.
\bb{11} P.G. de Gennes, ``Superconductivity in Metals and Alloys'', Benjamin,
New York (1966).
\bb{12} E. Capelutti and R. Zeyher, Phys. Rev. B {\bf 59}, 6475 (1999).
\bb{13} L. Benfatto, S. Caprara and C. Di Castro, Eur. Phys. J. B {\bf 17},
95 (2000).
\bb{14} S. Chakraverty, R.B Laughlin, D.K. Morr and C. Nayak, Phys. Rev B
{\bf 63}, 094503 (2001).
\bb{15} R. Zeyher and A. Greco, Phys. Stat. Sol. (b) {\bf 242}, 356 (2005).
\bb{16} Z.A. Hu, N.P. Ong, Y. Wang, T. Kakeshita and S. Uchida, Nature
{\bf 87}, 486 (1998).
\bb{17} Y. Wang, Z.A. Xu, T. Kakeshita, S. Uchida, S. Ono, Y. Ando and N.P. 
Ong, Phys. Rev. B {\bf 64}, 224519 (2001).
\bb{18} Y. Wang, N.P. Ong, Z.A. Xu, T. Kakeshita, S. Uchida, D.A. Bonn,
R. Liang and W.N. Hardy, Phys. Rev Lett. {\bf 87}, 257003 (2002).
\bb{19} C. Capan, K. Behnia, J. Hinderer, A.G.M. Jansen, W. Lang, C. Marcenat,
C. Marin and J. Flouquet, Phys. Rev. Lett. {\bf 88}, 056601 (2003).
\bb{20} K. Maki, B. Dora, A. Vanyolos and A. Virosztek, Curr. Appl. Phys.
{\bf 4}, 693 (2004).
\bb{21} V. Sandu, E. Cimpoiasu, T. Katuwai, Shi Li, 
M.B. Maple and C.C. Almasan, Phys. Rev. Lett. {\bf 93}, 177005 (2004).
\bb{22} B. Dora, K. Maki, A. Virosztek, Europhys. Lett. {\bf 72}, 1 (2005).
\bb{23} H. Won, S. Haas, D. Parker and K. Maki, Phys. Stat. Sol. (b)
{\bf 242}, 363 (2005).
\bb{24} H. Won, S. Haas, K. Maki, D. Parker, B. Dora and A. Virosztek,
cond-mat/0508234.
\bb{25} G.E. Volovik, JETP Lett. {\bf 58}, 496 (1993).
\bb{26} K.A. Moler et al, Phys. Rev. Lett. {\bf 73}, 2744 (1994).
\bb{27} B. Revaz et al, Phys. Rev. Lett. {\bf 80}, 3364 (1998).
\bb{28} S.J. Chen, C.F. Chang, H.L. Tsay, H.D. Yang and J.-Y. Lin
Phys. Rev. B {\bf 58}, 14753(R), (1998).
\bb{29} S. Nishizaki, Y. Maeno and Z. Mao, J. Phys. Soc. Jpn. {\bf 69},
573 (2000).
\bb{30} H. Won and K. Maki, Europhys. Lett. {\bf 52}, 427 (2000).
\bb{31} C. K\"ubert and P. Hirschfeld, Solid Stat. Comm. {\bf 105},
459 (1998).
\bb{32}C. K\"ubert and J.P. Hirschfeld, Phys. Rev. Lett. {\bf 80}, 4963 (1998)
\bb{33} I. Vehkter, J.P. Carbotte, E.J. Nicol, Phys. Rev. B 
{\bf 59}, 7123 (1999). 
\bb{35} H. Won and K. Maki, cond-mat/0004105.
\bb{36} T. Dahm, K. Maki and H. Won, cond-mat/0006301.
\bb{sal} H. Won, S. Haas, D. Parker, S. Telang, A. V\`anyolos and K. Maki,
in Lectures on the Physics of Highly Correlated Electron Systems IX, AIP 
proceedings 789 (Melville 2005), pp. 3-43.
\bb{37} K. Izawa, H. Takahashi, H. Yamaguchi, Yuji Matsuda, M. Suzuki, T. Sasaki, T. Fukase, Y. Yoshida,
R. Settai and Y. Onuki, Phys. Rev. Lett. {\bf 86}, 2653 (2001).
\bibitem{38} K. Izawa, H. Yamaguchi, Yuji Matsuda, H. Shishido, R. Settai and Y. Onuki,
Phys. Rev. Lett. {\bf 87}, 57002 (2001).
\bibitem{39} K. Izawa, H. Yamaguchi, T. Sasaki and Yuji Matsuda, Phys. Rev. Lett. {\bf 88}, 027002 (2002).
\bibitem{40} K. Izawa, K. Kamata, Y. Nakajima, Y. Matsuda, T. Watanabe, M. Nohara, H. Takagi,
P. Thalmeier and K. Maki, Phys. Rev. Lett. {\bf 89}, 137006 (2002).
\bibitem{41} K. Izawa, Y. Nakajima, J. Goryo, Y. Matsuda, S. Osaki, H. Sugawara, H. Sato, P. Thalmeier and
K. Maki, Phys. Rev. Lett. {\bf 90}, 117001 (2003).
\bb{42} K. Maki, S. Haas, D. Parker, H. Won, K. Izawa and Y. Matsuda,
Europhys. Lett. {\bf 65}, 720 (2004).
\bb{43} T. Watanabe et al, Phys. Rev. B {\bf 70}, 184502 (2004).
\bb{44} H. Won, D. Parker, K. Maki, T. Watanabe, K. Izawa and Y. Matsuda,
Phys. Rev. B {\bf 70}, 140509 (2004).
\bb{60} G. Sarma, J. Phys. Chem. Solids {\bf 24}, 1629 (1963).
\bb{61} H. Won, H. Jang and K. Maki, cond-mat/9901252
\bb{62} K. Maki and H. Won, Czech J. Phys. {\bf 46}, 1033 (1996);
Physica B {\bf 322}, 315 (2002).
\bb{45} I. Affleck and J. B. Marston, Phys. Rev. B {\bf 37}, 3744 (1988).
\bb{46} C. Wu, S. Capponi, S.C. Zhang, Phys. Rev. B {\bf 70}, 220505(R), 
(2004).
\bb{47} P. Thalmeier, Z. Phys. B {\bf 95}, 39 (1994).
\bb{48} J.C. Campuzano, H. DIng, M.R. Norman, M. Randeira, Physica B 
{\bf 259-261}, 517 (1999).
\bb{53} B. Dora, K. Maki and A. Virosztek, Mod. Phys. Lett. B {\bf 18}, 327 
(2004).
\bb{49} A.A. Nersesyan and G.I. Vachnadze, J. Low Temp. Phys. {\bf 77}, 
293 (1989).
\bb{50} A.A. Nersesyan, G.I. Japaridze and I.G. Kimeridze, J. Phys. Cond.
Matt. {\bf 3}, 3353 (1991).
\bb{51} K. Maki, B. Dora, M. Kartsovnik, A. Virosztek, B. Korin-Hamzi\'c
and M. Basleti\'c, Phys. Rev. Lett. {\bf 90}, 256402 (2003).
\bb{52} B. Dora, K. Maki, A. V\`anyolos
and A. Virosztek, Europhys. Lett.
{\bf 67}, 1024 (2004).
\bb{57} W. Kang, H.Y. Kang, Y.J. Jo and S. Uji, Synth. Metals {\bf 133-134},
13 (2003).
\bb{55} T. Hu, M. Xiao, V. Sandu, C.C. Almason, K. Maki, B. Dora, T.A. Sayles
and M.B. Maple, preprint.
\bb{66} M. Pinteri\'c, S. Tomi\'c and K. Maki, Physica C {\bf 408-410}, 75
(2004).
\bb{67} I.J. Lee, S.E. Brown, W. Yu, M.J. Naughton and P. Chaikin, Phys. Rev.
Lett. {\bf 94}, 197001 (2005).

\end{thebibliography}
\end{document}